\documentclass[times,10pt,twocolumn]{article}
\usepackage{latex8}
\usepackage{times}
\usepackage{graphicx}

\long\def\comment#1{}

\newcommand{\denselist}{
      \setlength{\itemsep}{0pt}
      \setlength{\parsep}{1.5pt}
      \setlength{\topsep}{1.5pt}
      \setlength{\parskip}{2pt}
      \setlength{\partopsep}{0pt}
      \setlength{\labelwidth}{1em}
      \setlength{\labelsep}{0.5em} }

\newcommand{\bdesc}{\begin{description}\denselist}
\newcommand{\edesc}{\end{description}}

\newcommand{\secref}[1]{Section~\ref{#1}}
\newcommand{\figref}[1]{Figure~\ref{#1}}

\begin{document}

\title{User Participation in Social Media: Digg Study}

\author{Kristina Lerman\\
USC Information Sciences Institute\\
4676 Admiralty Way\\
Marina del Rey, California 90292\\
lerman@isi.edu\\
}

\maketitle
\thispagestyle{empty}

\begin{abstract}
The social news aggregator Digg allows users to submit and moderate
stories by voting on (digging) them. As is true of most social
sites, user participation on Digg is non-uniformly distributed, with few users
contributing a disproportionate fraction of content.
We studied user
participation on Digg, to see whether it is motivated by
competition, fueled by user ranking, or social factors, such as
community acceptance.
 For our study we collected activity data of the top users weekly over the course of a year.
We computed the number of stories users
submitted, dugg or commented on weekly. We report a spike in user activity in
September 2006, followed by a gradual decline, which seems
unaffected by the elimination of user ranking. The spike can be
explained by a controversy that broke out at the
beginning of September 2006.
We believe  that the lasting acrimony that this incident has
created led to a decline of top user participation on Digg.

\end{abstract}

\section{Introduction}
Digg, which launched in 2004, is arguably one of the most popular and active of social news
sites. Digg's functionality is exceedingly simple: users submit links to
stories they find online, and other users rate them by voting on
them. Each day Digg selects a handful of stories to feature on its
front page. Although the exact formula for how a story is selected
for the front page is secret, so as to prevent users from
``gaming the system'' to promote advertising or spam, it appears to
take into account the number of votes a story receives~\cite{Lerman07ic}. The
front page, therefore, does not depend on the decisions of
a few editors, but emerges from the opinions of large number of users. This
type of collective decision making, called `wisdom of crowds', can be extremely
effective, outperforming special-purpose algorithms~\cite{Rose2006Rumsfeld}.

As of the writing of this paper, Digg has well over one million registered
users and more than 2,000 stories submitted daily. When a story
makes it to the front page, it generates thousands of views.
Digg's popularity has not escaped notice of
advertisers and marketers, who tried to exploit its popularity to
drive traffic to their sites. Digg continued to defend itself from
manipulation, by changing the algorithm it uses to promote
stories~\cite{diggblog}.

One recent victim of change was the Top Users list. Digg ranked users
according to how successful they were in getting their stories promoted to the front
page. Clicking on the Top
Users link allowed one to browse through the ranked list of users,
where \#1 user had the most front page stories, \#2 the second
most, etc.
There was speculation that ranking increased competition, leading
some users to be more active in submitting and digging stories on
the site in order to improve their
rank~\cite{roseTopUsers}. In February 2007, Digg discontinued making the Top Users list public,
citing concerns that marketers were paying top users to
promote their products and services~\cite{WSJ}. Currently, an unofficial Top Users list is
available through a third party.

We followed user activity on Digg over the course of about a year,
tracking the number of stories the users submitted, voted and
commented on, as well as their rankings. This long term view allows
us to examine the incentives that drive user participation in social
media. For example, does elimination of the Top Users list affect
user activity on Digg? Or does community acceptance encourage user participation~\cite{Joyce06,Qazvinian07}?
These questions have relevance to other
social media sites that operate on principles of social
participation similar to Diggs: Wikipedia, Flickr, and
others.

\section{Digg study}
For our study, we collected data by
scraping Digg with the help of Web wrappers.\footnote{ Wrappers were created using
tools provided by Fetch Technologies (http://fetch.com/).}
We trained the wrapper to extract information about the top 1000
recently active users from the Top Users list.  For each user, the wrapper extracted the number of stories
  submitted, commented and voted on (dugg); the number of stories
promoted to
  the front page; users's rank; the list of
  friends, and reverse friends (``people who
  have befriended this user''). The wrapper was executed weekly,
  starting in July 2006, until Digg stopped providing the Top Users list
  in February 2007 (although there were several weeks
  during which the wrapper was not working due to changes in site
  design). In April 2007, Digg made public an API to facilitate programmatic access to
its data. We used the API to retrieve data about the activities of the top 1,000
users, whose names came from the third-party Top Users
list\footnote{http://www.efinke.com/digg/topusers.html}.
The number of comments made by
users were not available through the API.
In all, we had 50 weeks of data
covering periods of Digg's phenomenal growth, as well as the controversies that
engulfed it.

\subsection{User activity}
\label{sec:activity}
\begin{figure*}[tbh]
\begin{tabular}{ccc}
  \includegraphics[height= 1.45in]{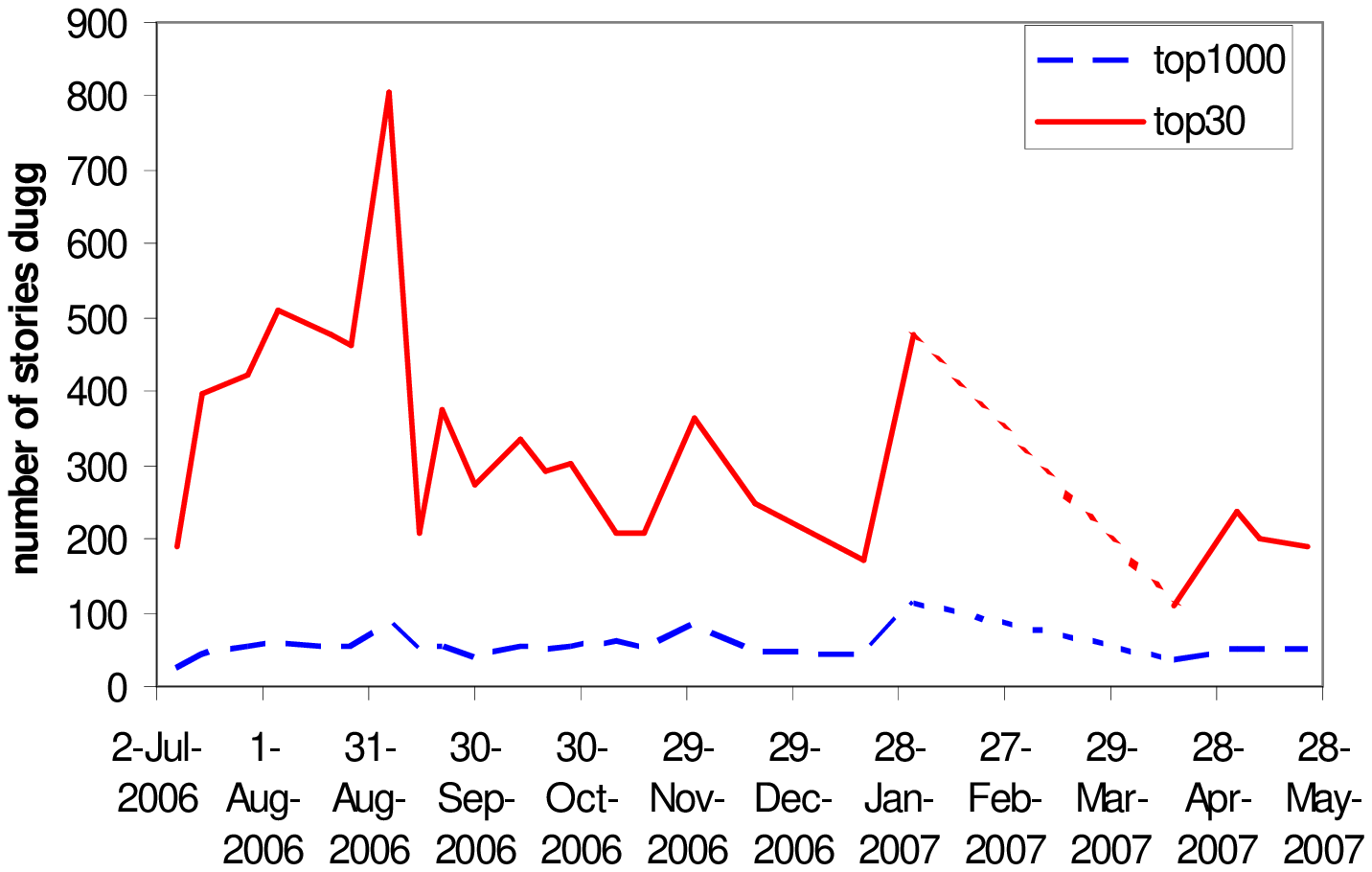} &
  \includegraphics[height= 1.45in]{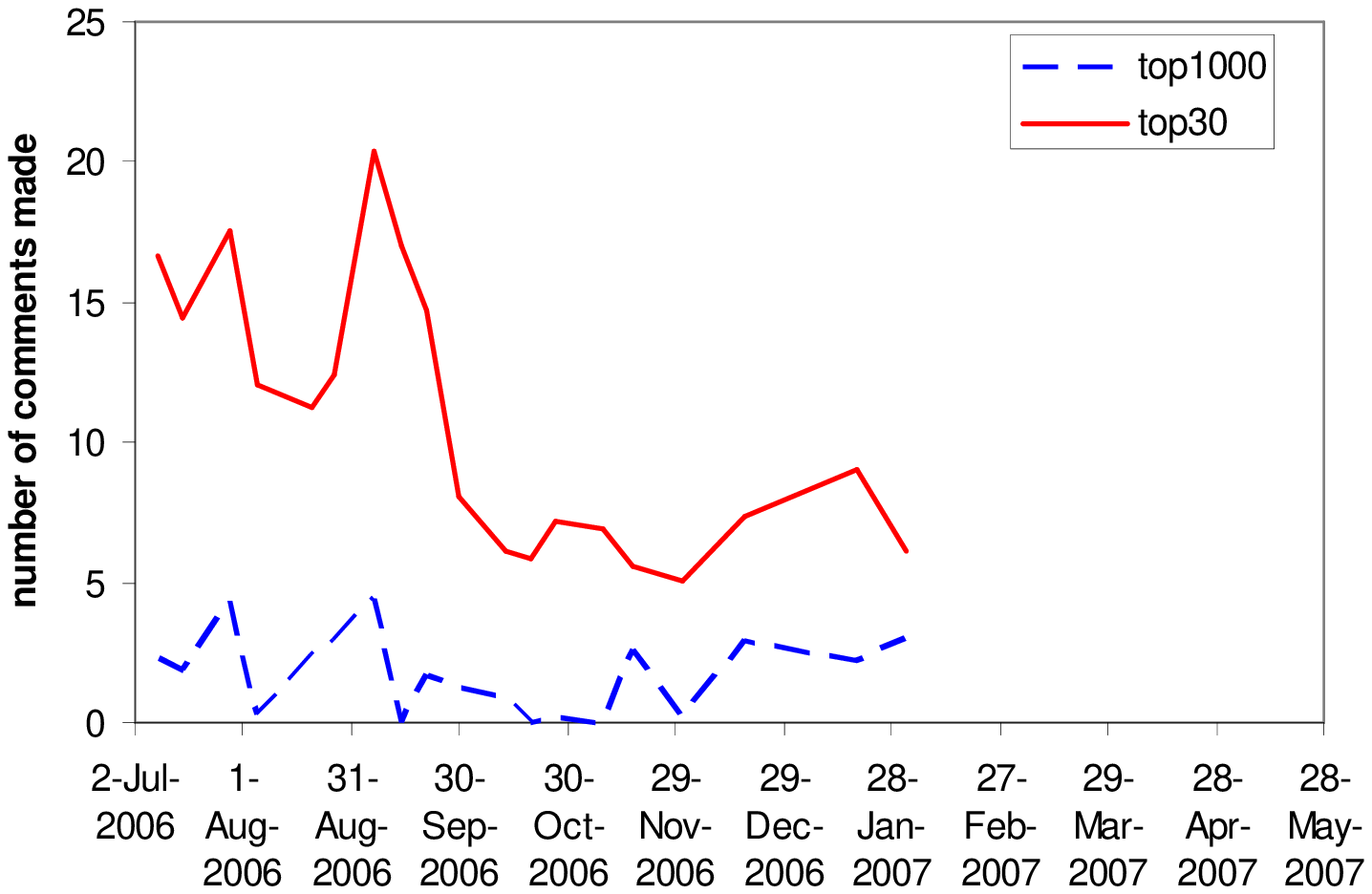} &
  \includegraphics[height= 1.45in]{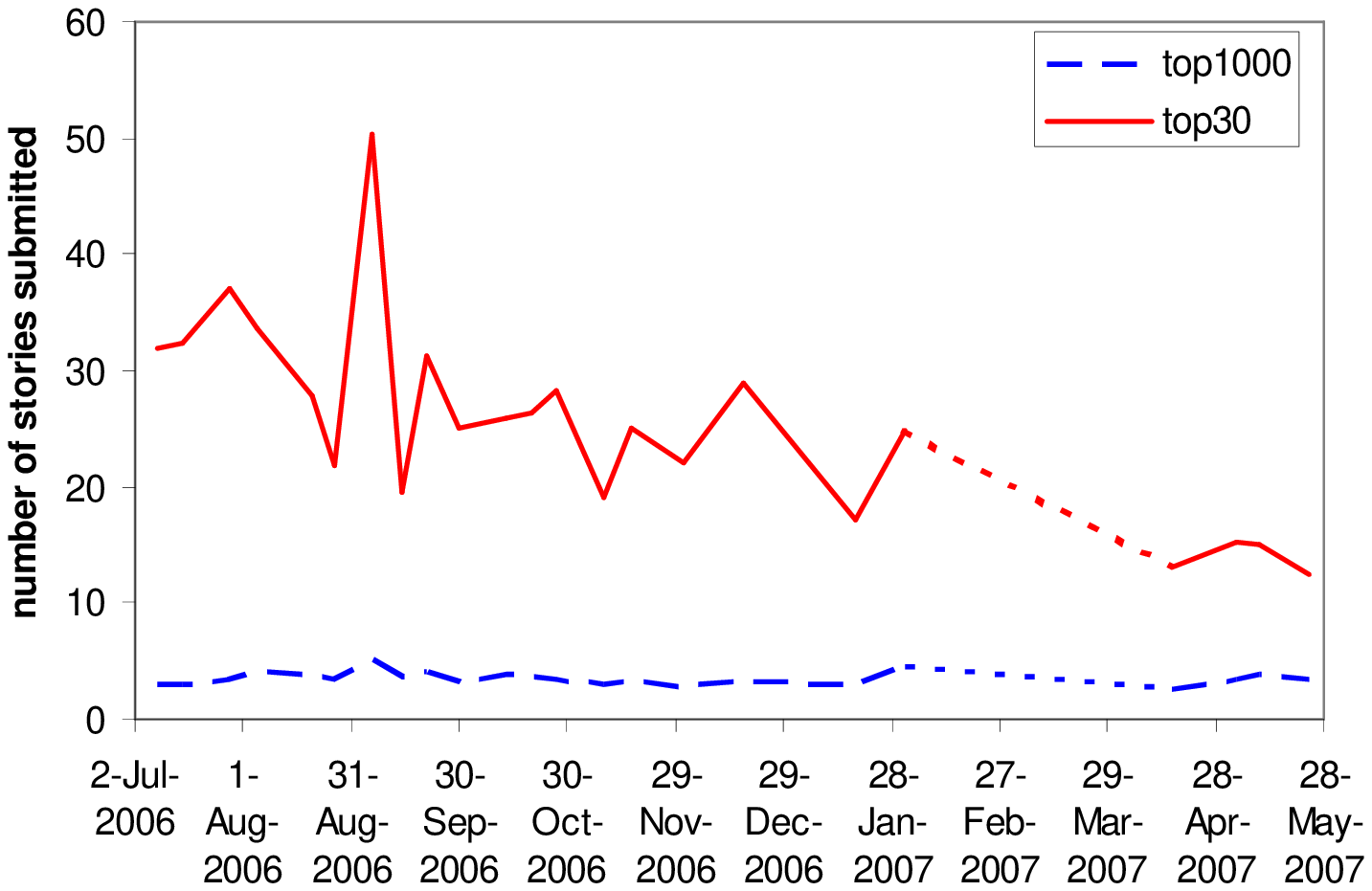} \\
  (a) & (b) & (c)
  \end{tabular}\caption{The average weekly (a) digging, (b) commenting, and (c) story submission rate for
  Digg's top1000 and top30 users} \label{fig:activity}
\end{figure*}

\figref{fig:activity} shows the average (per user) weekly activity on Digg.
We report separately the activity of the top 30 users (red)
and the top 1000 users (blue).
\figref{fig:activity} shows (a) the average weekly number of dugg stories, (b) the average weekly number
of comments, and (c) the average weekly number of new
submissions. The figure confirms that
top users are much more active than the lower-ranked users ---
contributing several times more stories, votes and comments.
The start of the dotted lines indicates discontinuation of the Top Users
list.

\begin{figure*}[tbh]
\begin{tabular}{cc}
 \includegraphics[height= 1.35in]{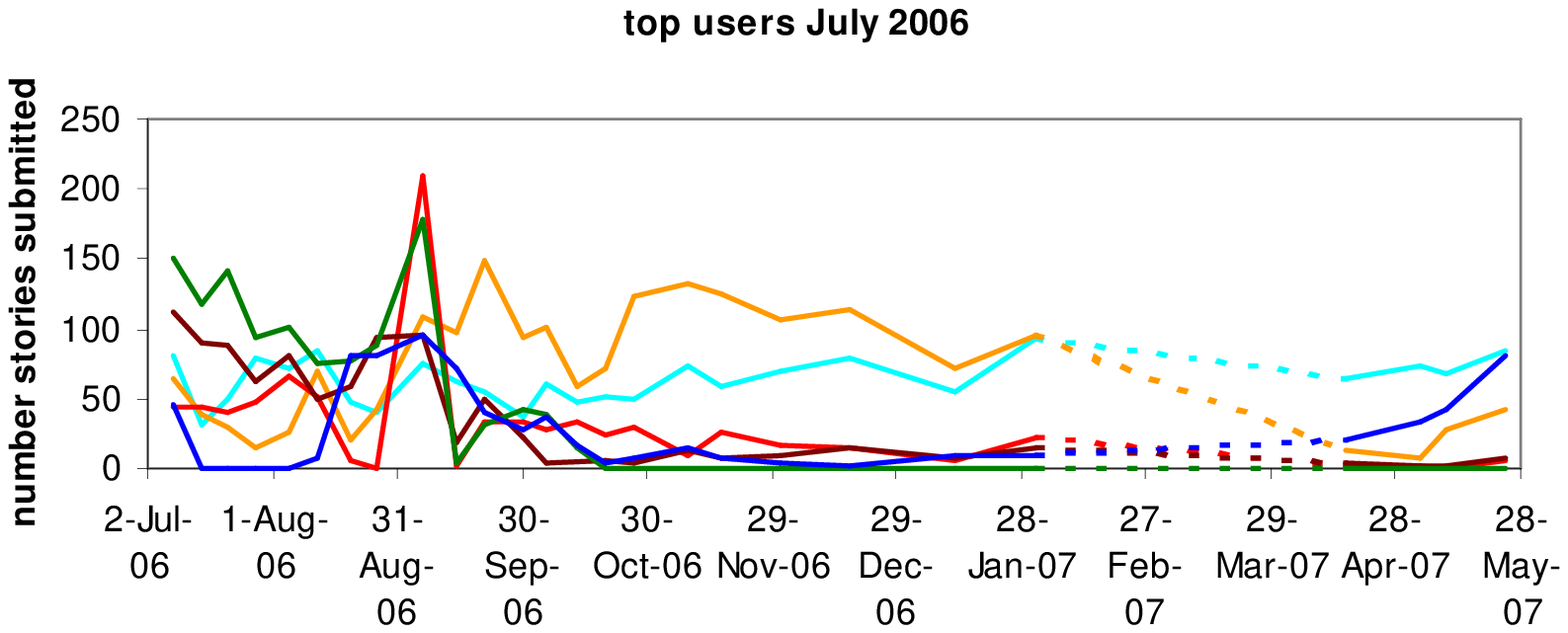} &
  \includegraphics[height= 1.35in]{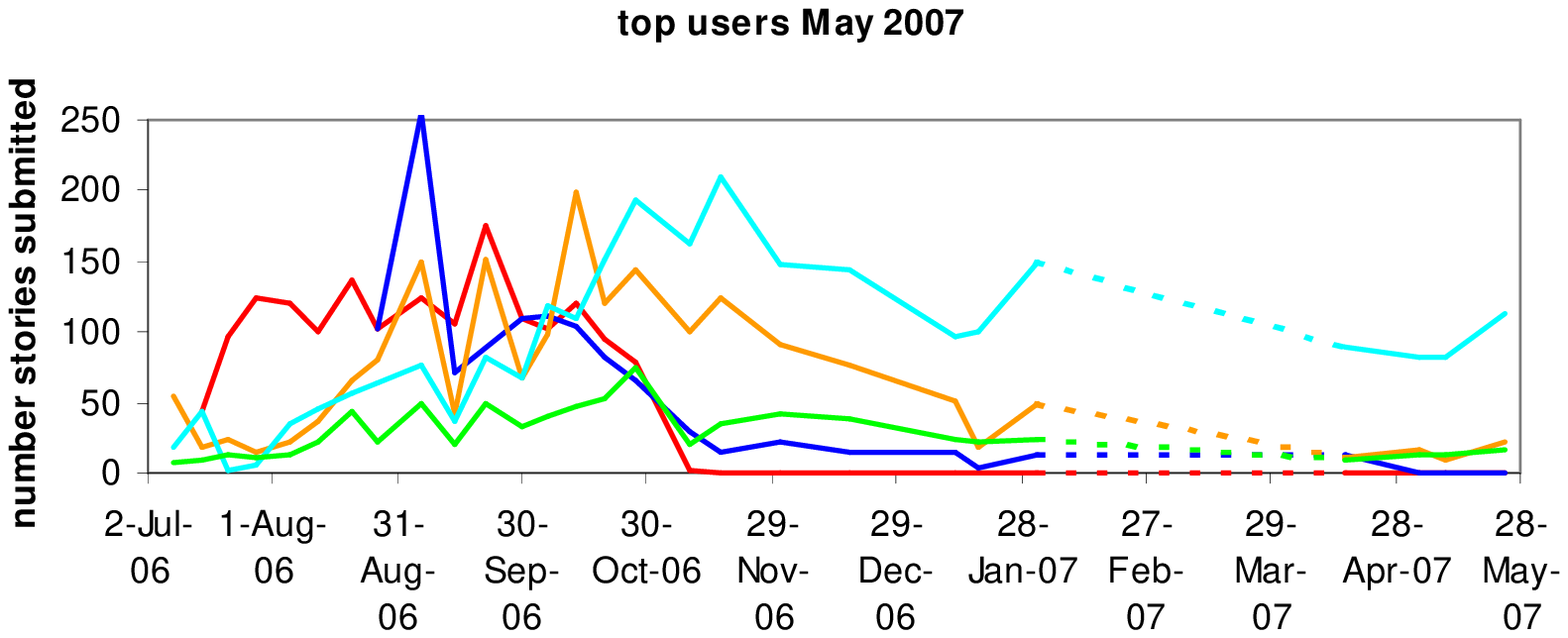} \\
 (a) & (b)  \\
  \includegraphics[height= 1.35in]{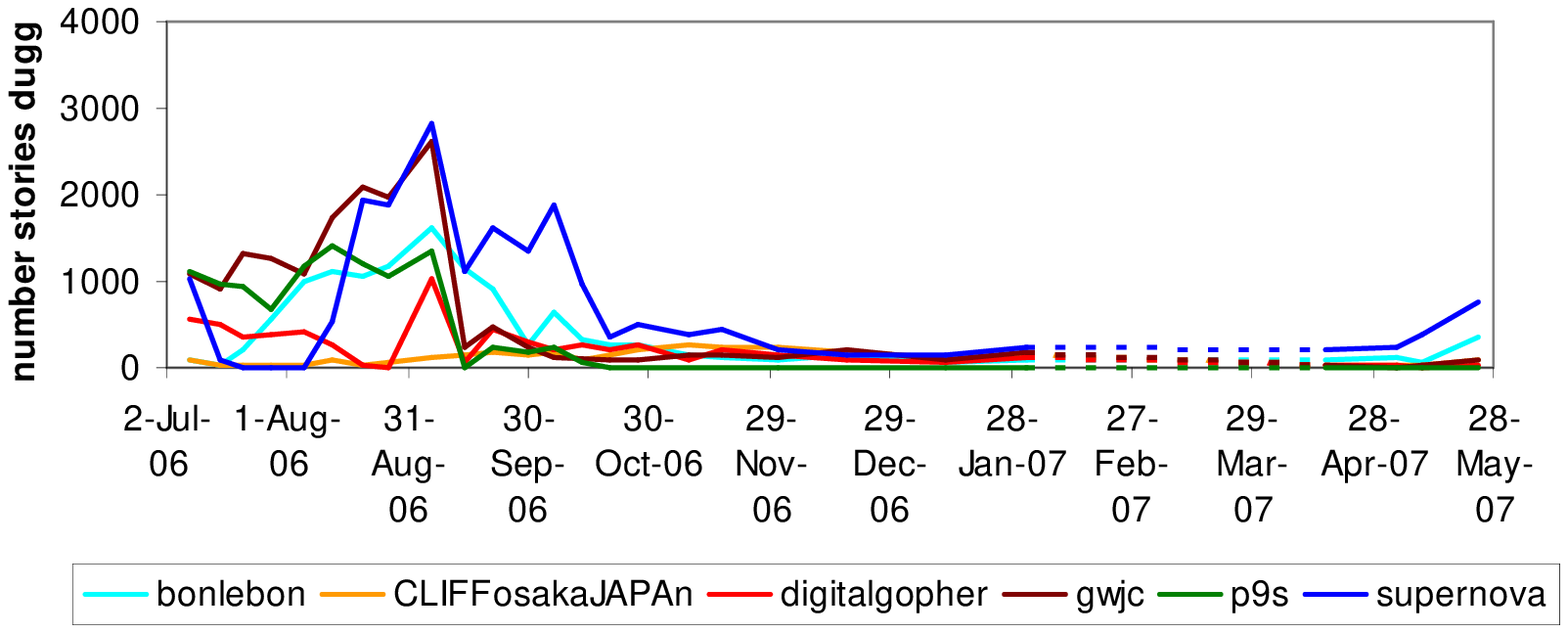} &
  \includegraphics[height= 1.35in]{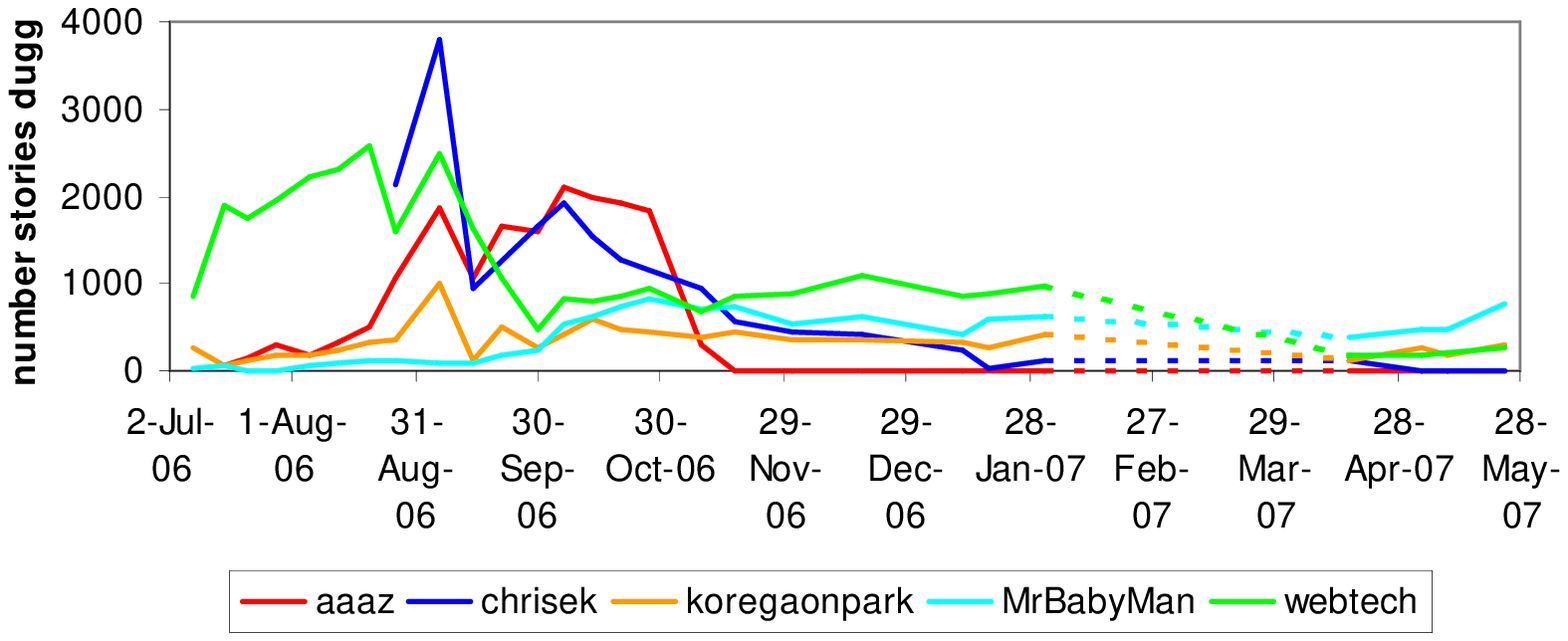} \\
 (c) & (d)

  \end{tabular}\caption{(a,c) Activity of users who were among the top users in July 2006 and
  still active in May 2007, and (b,d) newly active users who were
  in  the top 10 in May 2007. Plots (a,b) how the
  average weekly number of new submissions;
  (c,d) show the average weekly number of dugg stories. 
   } \label{fig:topactivity}
\end{figure*}

Did eliminating the Top Users list lead to a decrease in user
activity? Although the post-2/2007  digging and submission rates are smaller
for the top30 users,
they are still within the range of their pre-2/2007 levels. The
decline in participation (story submission, digging and
commenting rates) had started some weeks before
that. Considering that the overall number of new daily submissions
has been going up, the decline in
activity of top users is more than compensated by the increase in
the number of users.

The patterns seen in \figref{fig:activity} --- a spike in user activity at the
beginning of September 2006, followed by a gradual decline --- is
even more pronounced in the activities of select top10 users, shown in
\figref{fig:topactivity}. \figref{fig:topactivity}(a) \& (c) show the
activity (stories submitted and dugg) of ``old timers,'' or top10 users in July 2006
who were still ranked a year later.
\figref{fig:topactivity}(b) \& (d) show the activity of the ``newcomers,''
users who were not ranked in 2006, but attained the top10
status by May 2007.
Although a handful of users continued to submit new stories at the
same rate, there is a marked decrease in the activity
of both ``old timers'' and ``newcomers'' after September 2006. The decline is
greatest in the digging rate of ``old timers,'' followed by a
decline in the digging rate of ``newcomers.''
The submission rate is not as strongly affected by the
controversy as the digging rate, although there is also a decline in
this mode of participation post 9/2006. The elimination of
the Top Users list does not seem to have significantly affected the activity of
these users.

\subsection{The controversy}
\label{sec:controversy}
It is clear from the figures above that a dramatic event took place
at the beginning of September 2006, that had long lasting and
profound impact on user activity on Digg.
On September 5, 2006, a user posted an
analysis of the user activity statistics that,
similar to our findings, showed that the top 30 users were
responsible for a disproportionate fraction of the front page stories.\footnote{http://jesusphreak.infogami.com/blog/is\_digg\_rigged}
This analysis meant to support the claim that
top users conspired to automatically promote each other's stories, or
as a blogger Michael Arrington put the next day,
``a small group of powerful Digg users, acting together, control a large percentage of total home page
stories''~\cite{techcrunch}. Needless to say, these accusations
incensed both sides: the general Digg population, who felt that
Digg's democratic ideal was compromised by a 'cabal' of top users,
and the top users, who received the brunt of the anger. The
escalating war of words was fought on blogs, Digg's pages (as
evidenced by the spike in activity in early September 2006),
and it even attracted the attention of mainstream media~\cite{USAToday}.
Within days, Digg's management announced changes to the promotion algorithm
that devalued ``bloc voting'' or votes coming from
friends~\cite{diggblog}.\footnote{The new promotion algorithm, implemented in November 2006, appears to have been successful at
reducing the top user dominance of the front page.} Top users saw this as a repudiation of
their contributions to Digg, and at least one top user, who held the No. 1 position
at the time, publicly resigned~\cite{wired}.

\comment{
\begin{figure}[tbh]
  \includegraphics[width= 3in]{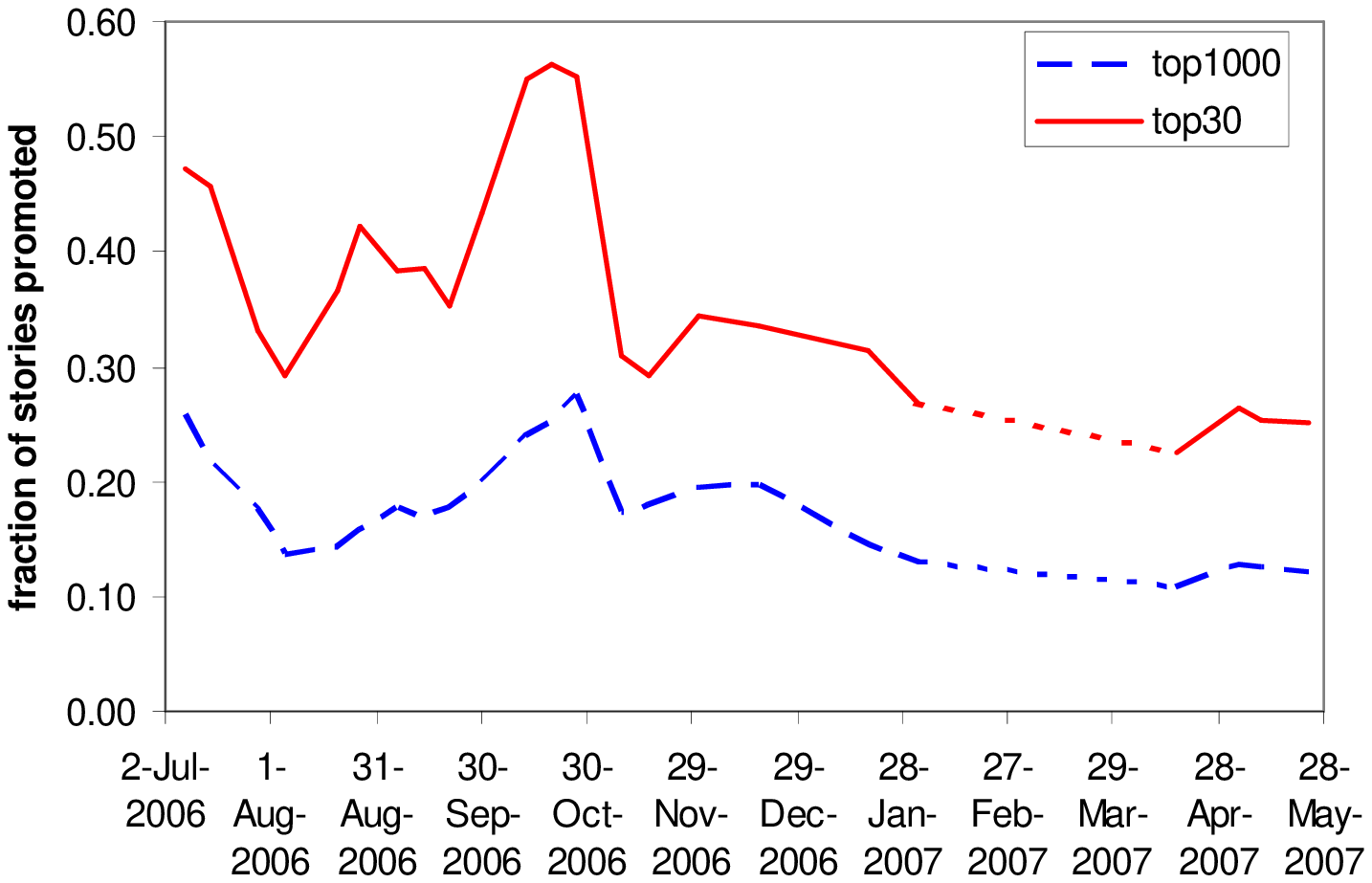} \\
 \caption{Average success rate per user} \label{fig:friends}
 \end{figure}
 }

Remarkably, the top users controversy did not seem to affect the
growth of social networks on Digg. The
average number of  new friends and  reverse friends (users who
befriended a particular user) added weekly by the for the top 30 and the top 1000 users
did not seem to be impacted by the controversy.
In fact, the week the controversy broke corresponds to a local peak
both in the number of new friends and reverse friends for both the
top 30 and the top 1000 users. Only two weeks to a month later do we
see evidence of users taking other users off their friends list.

\comment{
Remarkably, the top users controversy did not seem to affect the
growth of social networks on Digg. \figref{fig:friends}(b) shows the
average number of  new friends and  reverse friends (users who
befriended a particular user) for the top 30 and the top 1000 users.
In fact, the week the controversy broke corresponds to a local peak
both in the number of new friends and reverse friends for both the
top 30 and the top 1000 users. Only two weeks to a month later do we
see evidence of users taking other users off their friends list.

}

\subsection{Top 10 composition}
\label{sec:top10}

\begin{figure*}[tbh]
\begin{tabular}{cc}
  \includegraphics[height=1.2in]{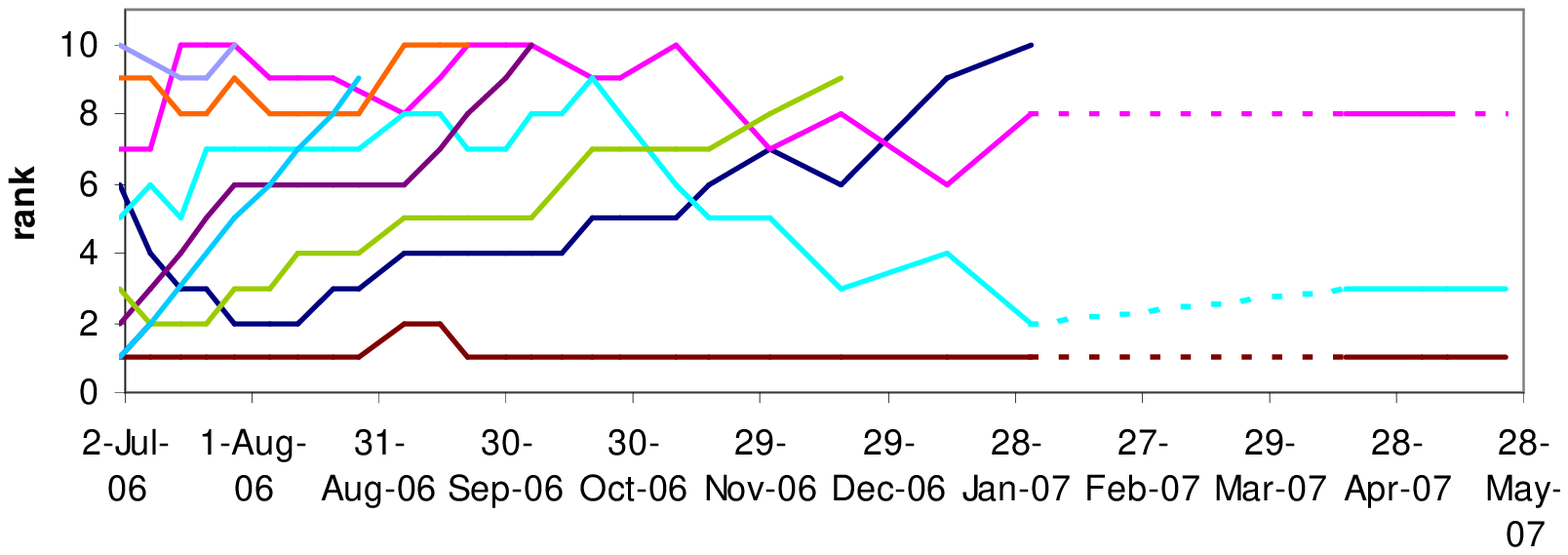} &
  \includegraphics[height=1.2in]{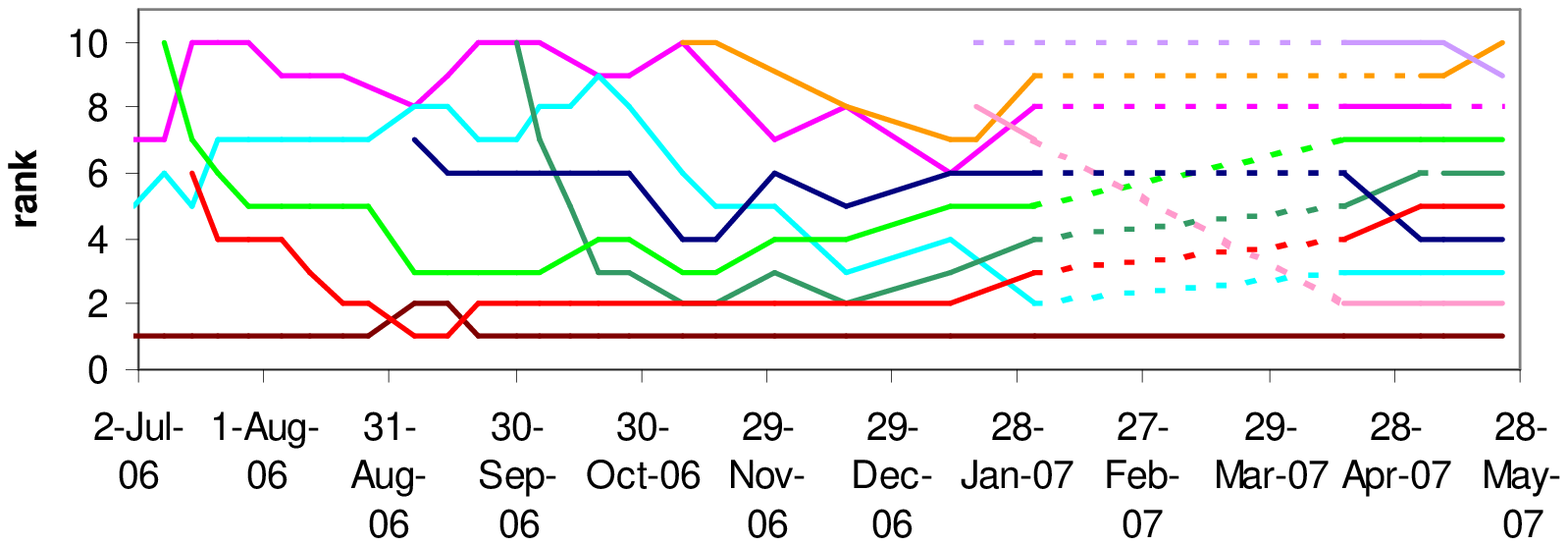} \\
  \end{tabular}\caption{Evolution of user rank of users who were (a)
  in the top10 in July 2006 and (b) May 2007. } \label{fig:rank}
\end{figure*}

\figref{fig:rank} shows evolution of rank for users who were in the top10 in
July 2006 and in May 2007. Only three of
the July 2006 top10 users retained their top10 status nearly a year later.
Two of these users, \emph{BloodJunkie} and \emph{p9s}, stopped contributing
altogether;\footnote{Another user, aaaz, who reached the top10 since July
2006 stopped participating in Digg shortly after reaching the top10.}
other users saw their rank slip due to decline in activity.
Prior to the elimination of the Top Users list, there was continual
turmoil in user positions within the Top Users list, but after its elimination,
the ranks did not change quite as much. This observation is not a byproduct of differences in post-2/2007 user
ranking, because the same effect is present if the users were ranked
simply by the number of the front page stories they have in
pre-2/2007 data.

\section{Discussion}

The Digg dataset allows us to study incentives to user participation on a social
media site. User participation in most, if
not all social media sites, is non-uniformly distributed, with a few users doing a
large fraction of the work, whether it is editing Wikipedia articles,
writing open source software, contributing videos (YouTube) or moderating news
stories (Digg). For example, in July 2006, the top 3\% of the top
1,000 users made 33\% of the weekly submissions,
21\% of the diggs and 60\% of the stories promoted to the front page.
This type of heavy-tail distribution has been expressed as
Pareto principle:  ``80\% of the work is done by
20\% of individuals.''
Keeping the top users
happy should be a priority of a social media site.

The Digg dataset allowed us to indirectly study incentives that influence
user participation: (a) competition, which manifests itself as a desire to improve one's standing in the
community, (b) social factors, such as community acceptance,
and (c) internal factors, e.g., user's success in getting his stories
promoted, which is affected by Digg's promotion algorithm.

According to Digg founder Kevin Rose, Digg first introduced the Top
Users list to encourage users to submit stories~\cite{roseTopUsers},
believing that the desire to improve one's position on the Top Users
list will motivate some users to devote significant portion of their
time to submitting and digging stories. If
this were true, then eliminating the Top Users list may lead
to a decrease in user activity. We did indeed see lower activity levels
after February
2007 (\secref{sec:activity}); however, this decline has been ongoing for weeks prior to
this date. The only tangible consequence we
observed was that user rank became more static (\secref{sec:top10}).

So why has the activity of top users declined? This could be
explained by two factors: internal changes in Digg's promotion
algorithm, which made it harder for top users to get their stories
promoted, and social factors. In September 2006
Digg promised a major change in its story promotion algorithm~\cite{diggblog},
which was implemented in November 2006. The new promotion algorithm
attempted to decrease the top user monopoly of the front page, and
it did lead to decrease of the success rate of the top30 users.
However, the drop in user participation, as seen in the drop in the
number of stories dugg and the number of comments made, was already
ongoing. This drop can be traced to September
2006, when a controversy broke on Digg
about the Top User ``conspiracy'' to control the front page.
Social recognition is the glue that holds the community together,
and is a more powerful motivator than competition. Recognition in social media comes
in the form of comments, votes on content one has
submitted, or friendship requests. Positive recognition motivates
the user to remain active or increase activity~\cite{Qazvinian07,Joyce06},
while negative recognition can destroy the community~\cite{trolls}.
We believe  that the lasting acrimony that this incident
created has led to a general decline in individual user participation on Digg. The
declined did not affect just the highest ranked users, but the rest
of the community as well. While decline in the activity of top users is
offset by rising membership, it is not clear what long term impact
on Digg the controversy will have.

\paragraph{Acknowledgement} This research is based on work supported
in part by the National Science Foundation under Award Nos.
IIS-0535182 and IIS-0413321. We are grateful to Fetch Technologies for providing
wrapper building and execution tools and to Dipsy Kapoor for computing statistics
from the extracted data.

\bibliographystyle{latex8}
\bibliography{../../social,../../lerman,../../robots}

\end{document}